\documentclass{kapproc}

\normallatexbib

\begin{document}
\input epsf 

\articletitle{At the Interface of Quantum and Gravitational Realms}

\author{D.\ V.\ Ahluwalia  }
\affil{Theoretical Physics Group,
Facultad de Fisica, UAZ, 
A.\ P.\ C-600, Zacatecas, ZAC 98062, Mexico}
\email{ahluwalia@phases.reduaz.mx; http://phases.reduaz.mx}

\chaptitlerunninghead{At the Interface of Quantum and gravitational Realms
 }

\begin{abstract}
In this talk I review a series of recent conceptual 
developments at the interface of the quantum and gravitational 
realms. Wherever possible, I comment on the 
possibility to probe the interface experimentally. 
It is concluded that the underlying spacetime for
a quantum theory of gravity must be non-commutative, that wave-particle
duality suffers significant modification at the Planck scale, and 
that the latter forbids probing spacetime below Planck length. Furthermore,
study of quantum test particles in classical and quantum sources of 
gravity puts forward theoretical challenges and new experimental
possibilities. It is suggested that existing technology may
allow to probe
gravitationally-modified wave particle duality in the laboratory,
\end{abstract}


\def\beq{\begin{eqnarray}}
\def\eeq{\end{eqnarray}}

\section*{Introduction}
The purpose of this written version of the talk given at the ``Mexican 
Meeting on Mathematical and Experimental Physics (Colegio Nacional, 
10-14 September 2001)'' is to briefly review some of the conceptual 
developments at the interface of the quantum and gravitational realms. 
The version presented here primarily confines to the contributions
I have made, and is by no means intended as a review of this developing
field. I am also happy, when possible, to point reader's attention  to 
the relevant experimental literature.

\section*{Non-commutative Nature of Spacetime and
Gravita\-tionally-Modified Wave Particle Duality}

\noindent
{\em Setting the Stage:
Quantum Measurement with Gravitational Effects Ignored.}

The experimental foundations of  quantum mechanics 
reside in the {\em in principle\/} lower limit on the extent
to which the unavoidable disturbance that 
the position and momentum measurements carry can be 
reduced.\footnote{This remark applies equally well to any set
of canonically conjugate variables.} 
This circumstance arises from the experimental implication of the 
photo-electric effect. It tells us that the energy carried by a  light beam 
is not a continuous variable. Its intensity can only be changed in discrete
units of $\hbar\omega$, where  $\omega$ is the angular frequency
of the probing light beam. This is encoded in  fundamental
commutators, such as:
\beq
\left[x,\,p_x\right]
=i\hbar\,,\quad \left[x,\,y
\right] =0\,.
\eeq
In configuration space, a solution of the above commutators
is:
\beq
p_x=\frac{\hbar}{i}\frac{\partial}{\partial x}\,.
\eeq
The operator $p_x$ carries with it non-renormalizable
eigenfunctions of the form 
\beq
\psi(x)\sim\exp\left({{i p_x x}\over{\hbar}}\right)\,.
\eeq
However, integrating over different-momenta eigenfunctions one
may obtain well-defined and normalizable wave packets that
describe space-localizable particles. Now let $\lambda$ be
the spatial periodicity of 
$\psi(x)$. That is, $x\to x\pm \lambda$, advances the phase
of $\psi(x)$ by $\pm 2\pi$.
Then, we find that a particle of momentum $p_x$ carries with a
de Broglie wave length:
\beq
\lambda_{dB}
= {{h}\over{\vert p_x\vert }}\,.
\eeq
In a commutative spacetime this lies at the heart of 
the quantum-mechanical wave particle duality.

\noindent
{\em Quantum Measurement with Gravitational Effects 
Incorporated.}

The usual measurement process in quantum mechanics
ignores any gravitational effects that may be inherent
in it. Such effects become important in quantum gravity 
because, in general, there is an {\em unavoidable\/} change 
in the energy-momentum tensor associated with
the collapse of a wave function.\footnote{Such a collapse 
may be associated with a position measurement, or
position measurements of different components of a position
vector.} Consequently, 
quantum measurements of spatio-temporal locations of events
cannot ignore inherent gravitational effects. These
effects, we hasten to add are intrinsic to the events
under consideration. In the first approximation, 
they do not directly refer to the background curvature 
\cite{grf1994,jm,gv,sh,as,gac,jn,mm,ns,fs,cls,KMN}.

When gravitational effects in a quantum measurement process
are incorporated it is found that there is an {\em in-principle and 
unavoidable\/} non-comm\-uta\-tivity of the position and temporal
measurements. Furthermore, the fundamental commutator undergoes a
change of the form \cite{KMN}:
\beq
\left[x,\,p_x\right]
= i\hbar\left[1+\epsilon \frac{\lambda_P^2 \,p_x^2}{\hbar^2}\right]\,,
\eeq
where
\beq
\lambda_P=
\sqrt{\frac{\hbar G}{c^3}}\,,\label{mod_funcomm}
\eeq
and $\epsilon$ is a number of the order of unity (to be set
to unity, now onwards). Given Eq. (\ref{mod_funcomm}),
the introductory discussion contained in 
``{\em Setting the Stage: 
Quantum Measurement with Gravitational Effects Ignored\/}''
immediately implies that wave particle duality must suffer
a fundamental change at the Planck scale. 
For one-dimensional motion, the change in wave-particle
duality takes the form  \cite{pla2000,KMN}:
\beq
\lambda=
 \frac{\overline{\lambda}_P}{\tan^{-1}(\overline\lambda_P/\lambda_{dB})}
\eeq
In the above equation we have defined, 
$\overline{\lambda}_P = 2\pi\lambda_P$.
It is readily seen that in the low-energy limit
the gravitationally modified wave length $\lambda$ reduces to
the well-known de Broglie wave length. In the Planck regime, however,
something surprising, and welcome, happens. 
The gravitationally modified wave length $\lambda$
saturates to (4 times, with $\epsilon=1$) 
the Planck length. A similar result was later
obtained by Bruno, Amelino-Camelia, and  Kowalski-Glikman 
\cite{gac_saturation} (Also, see important observations
of Padmanabhan in Refs. \cite{tp1,tp2}). 
Theoretically, this result may be interpreted as that no particle
wavelengths are available to probe spacetime below the Planck
length distances ($\lambda_P^2$ areas, and $\lambda_P^3$ volumes).   
Moreover, this saturation also suggests that in some sense 
the relativity, special and general, must suffer changes so that
length contractions below $\lambda_P$ do not occur. 

Theoretically, one is, therefore, called upon to develop
a relativity that carries not only an inertial-observer independent 
velocity, i.e. $c$, but also a similarly independent length
scale. The latter may be identified with $\epsilon \lambda_P$. 
Amelino-Camelia has already undertaken the task of building
such a modification to the relativity theory \cite{gac3}.

Experimentally, the derived saturation of $\lambda$, 
implies freezing of neutrino oscillations
at the Planck energies, and carries 
several  phenomenological implications \cite{pla2000}.
However, the phenomenological implications may not be
confined to early universe alone, as we now argue.

Superconducting quantum interference devices (SQUIDs), 
when cooled sufficiently below the
critical temperature, 
may carry temp\-erature-tunable superconducting currents  
with total superconducting mass 
\beq
m_s\sim f(T)\, N_a\, m_c\,,\label{ms}
\eeq
behaving as one quantum object (under certain circumstances).
In Eq. (\ref{ms}),  
$N_a \approx 6 \times 10^{23}\,\, \mbox{mole}^{-1}$, 
$m_c\approx 2\times 0.9 \times 10^{-27}$ gm, and $f(T)$
encodes fraction of the available electrons that are in
a superconducting Cooper state at temperature, $T$. 
Sufficiently below the critical
temp\-rat\-ure,   $f(T)$ may approach unity.
The temperature-tunable, $m_s$, can easily  
compete with Planck mass,
\beq
m_P =\sqrt{\frac{\hbar c}{G}} \approx 2.2\times 10^{-5} \,\,\mbox{gm}\,.
\eeq
Thus, 
SQUIDs carry significant potential to probe
wave-particle duality near the Planck scale. The theoretical 
and experimental problem that remains to be attended is
to devise an experiment that invokes   
$m_s$, and not $m_c$.

\section*{Quantum Test Particles in Classical Sources of Gravity}

Principle of equivalence with classical test particles in
classical source of gravity has been verified to a remarkable
accuracy, see, e.g.,  \cite{cl_ep}. In this section we devote our attention
to quantum test particles when the source of gravity is treated
as a classical background.

The quantum  behavior of a mass eigenstate in the classical
source of gravity is well studied in the pioneering experiments
on neutron interferometry \cite{prl_cow,recent_cow}. In recent
such experiments an apparent violation of the principle of 
equivalence seems evident at roughly a part in one thousand
\cite{recent_cow}. If this result is not due to a yet unknown 
systematic error, then quantum mechanical motion of neutron 
in classical source of gravity poses serious theoretical challenge 
for its understanding \cite{plb2000,grg2001}. 

In atomic interferometry \cite{chu} the principle of equivalence 
is confirmed to a few parts in $10^9$.

To study the possibilities that quantum test particles offer to
study gravitational field let's us first note that the local 
gravitational potential in the solar system carries two sources: 
(a) Solar-system sources, such as Earth, (b) Cosmological sources, 
such as the local super-cluster of galaxies.
The former, on the surface of Earth, when measured in dimensionless
units is, $-7 \times 10^{-10}$, 
and varies as $R_\oplus/(z+R_\oplus)$ --- where $z$ is the vertical distance
from the surface of the Earth, and $R_\oplus$ is Earth's radius.   
While the latter can be estimated to be roughly $3 \times 10^{-5}$ --- see, 
Ref. \cite{plb2000}. 
It is roughly constant over the solar system.

Potentials of the type ''a'' induce not only gravitational forces, 
but they also are also responsible for observable 
gravitationally-induced phas\-es and the accompanying quantum 
interference effects. The type-b potentials  are essentially 
force free, and they have the net effect of red-shifting local clocks.
However, an experiment that seeks to study a possible violation
of equivalence principle must treat such potentials with due
care. In particular, given a violation of equivalence principle,
the type-b potentials acquire a {\em local\/} observability.
Quantum system -- modeled after flavor-oscillation clocks 
\cite{grf1996,grf1997,prd1998,kk1998,jw2001,gl,ac} --- appear to
be most sensitive experimental probes for type-b potentials.
There is now a significant and growing literature on
the subject of flavor-oscillation clocks and I refer to the just
cited list of references for the involved details.

\section*{Quantum Test Particles in Quantum Sources of Gravity}

The next level of theoretical and experimental sophistication
is called upon when one treats both the sources and the
test particles as quantum objects. The example of the former
is provided by a SQUID in a linear superposition of two counter-propagating
super-currents \cite{cpsc}. An example of the latter is once again
a system of flavor-oscillations clocks.

In the absence of a complete theory of quantum gravity, 
it is a non-trivial theoretical task to model the gravitational 
field of a quantum source of gravity. Yet, in the weak field,
non-relativistic, regime the gravitational field may simply
be taken as a quantum linear superposition of configurations
with classical counterparts. To our knowledge, 
not even a preliminary analysis 
exists on the subject. However, as is apparent, such a theoretical
undertaking is likely to prove a fruitful playing ground on the
interface of gravitational and quantum realms.   

\section*{Spatial and Temporal Fluctuations in Spacetime Foam}

Amelino-Camelia has argued that {\em spatial\/} fluctuations 
of the space-time foam can be experimentally probed in
gravity wave interferometers. This is a new and unexpected 
observation and may provide direct evidence for
quantum-gravity induced effects \cite{gac2,dva_nv}.

Complementing Amelino-Camelia's work, Kirchbach and I have put 
forward a thesis that the observed matter-antimatter asymmetry 
may arise from asymmetric space-time fluctuations and their 
interplay with the St\"uckelberg-Feynman interpretation of 
antimatter \cite{grf2001}. The thesis also argues that the effect 
of spacetime
fluctuations is to diminish the fine structure constant,
$\alpha=e^2/\hbar c$, in the past. Recent studies of the QSO absorption
lines provide a $4.1$ standard deviation support for this prediction 
\cite{QSO1,QSO2,QSO3}.
It is entirely possible that the empirical data on the fine structure
constant has already detected first signatures of the quantum-gravity
induced spacetime fluctuations.

\section*{Concluding Remarks
}

Towards building a quantum theory
of gravity, the interface of the gravitational 
and quantum realms is a rich
conceptual and experimental arena.
Here, theorists and experimentalists
alike may play with much profit.  
In this talk I have outlined mostly my personal 
contributions. It is apparent that the underlying spacetime for
a quantum theory of gravity must be non-commutative, that wave-particle
duality suffers significant modification at the Planck scale, and 
that the latter forbids probing spacetime below Planck length. Furthermore,
study of quantum test particles in classical and quantum sources of 
gravity puts forward theoretical challenges and new experimental
possibilities. 

\vspace{1cm}

It is my pleasure to thank A. Macias for an invitation to
present these results, and for arranging a well-attended conference
in the stimulating environment of Colegio Nacional. I also thank
T. Padmanabhan and D. Sudarsky
for several stimulating discussions on the subject,
and extend my apologies for not being able to track down
several of their relevant publications under the tight submission
deadline for this manuscript. 

\noindent
{\em This work is being supported by CONACyT project No 32067-E.}


\begin{chapthebibliography}{1}

\bibitem{grf1994}
D.\ V.\ Ahluwalia,
{\em Phys. Lett. B\/} {\bf 339} (1994) 301.

\bibitem{jm}
J.\ Madore, gr-qc/9709002.

\bibitem{gv}
G.\ Veneziano, 
{\em Europhys. Lett.\/} {\bf 2} (1986) 199.

\bibitem{sh}
S.\ de\ Haro,
{\em Class. Quant. Grav.\/} {\bf 15} (1998) 519.

\bibitem{as}
R.\ J.\ Adler, D.\ I.\ Santiago,
{\em Mod. Phys. Lett. A\/} {\bf 14} (1999) 1371.

\bibitem{gac}
G. Amelino-Camelia,
{\em Mod. Phys. Lett. A\/} {\bf 12} (1997) 1387.

\bibitem{jn}  
J.\ Y.\ Ng, gr-qc/0201022.



\bibitem{mm}
M.\ Maggiore,
{\em Phys. Lett. B\/} {\bf 304} (1993)   65.

\bibitem{ns}
N.\ Sasakura,  
{\em Prog. Theor. Phys.\/} {\bf 102 } (1999)   169.

\bibitem{fs}  
F.\ Scardigli
{\em Phys. Lett. B\/} {\bf 452} (1999)   39.


\bibitem{cls}
S.\ Capozziello, G.\ Lambiase, G.\ Scarpetta,  
{\em Int. J. Theor. Phys.\/} {\bf 39} (2000)   15.


\bibitem{KMN}
A.\ Kempf, G.\ Mangano, R.\ B.\ Mann,  
{\em Phys. Rev. D\/} {\bf 52} (1995)   1108.


\bibitem{pla2000} 
D.\ V.\ Ahluwalia, 
{\em Phys. Lett. A\/} {\bf 275 } (2000)   31.

\bibitem{gac_saturation}  
N.\ R.\ Bruno, G.\ Amelino-Camelia, J.\ Kowalski-Glikman
{\em Phys. Lett. B\/} {\bf 522 } (2001)   133.



\bibitem{tp1}  
T.\ Padmanabhan
{\em  Phys. Rev. Lett.\/} {\bf 78} (1997)   1854.

\bibitem{tp2}  
T.\ Padmanabhan, 
{\em Class. Quant. Grav.\/}
{\bf 4} (1987) L107.


\bibitem{gac3}
G. Amelino-Camelia,
{\em Int. J. Mod. Phys. D\/} {\bf 11} (2002) 35.


\bibitem{cl_ep}
G.\ L.\ Smith {\em et al.,\/}  
{\em Phys. Rev. D\/} {\bf 61} (2000)   022001.

\bibitem{prl_cow}  
R.\ Colella, A.\ W.\ Overhauser, S.\ A.\ Werner,
{\em Phys. Rev. Lett.\/} {\bf 34} (1975)   1472.




\bibitem{recent_cow}  
K.\ C.\ Littrel, B.\ E.\ Allman, S.\ A.\ Werner,
{\em Phys. Rev. A\/} {\bf 56} (1997)   1767.



\bibitem{plb2000}  
G.\ Z.\ Adunas, E.\ Rodriguez-Milla, D.\ V.\ Ahluwalia,
{\em Phys. Lett. B\/} {\bf 485} (2000)   215.

\bibitem{grg2001}  
G.\ Z.\ Adunas, E.\ Rodriguez-Milla, D.\ V.\ Ahluwalia,
{\em Gen. Rel. Grav.\/} {\bf 33} (2001)   183.


\bibitem{chu}  
A.\ Peters, K.\ Y.\ Chung, S.\ Chu,
{\em Nature\/} {\bf 400} (1999)   849.

\bibitem{grf1996}  
D.\ V.\ Ahluwalia, C.\ Burgard, 
{\em Gen. Rel. Grav.\/} {\bf 28} (1996)   1161. Erratum
 {\bf 29} (1997)   681. 

\bibitem{grf1997}  
D.\ V.\ Ahluwalia,
{\em Gen. Rel. Grav.\/} {\bf 29} (1997)   1491.

\bibitem{prd1998}  
D.\ V.\ Ahluwalia, C.\ Burgard, 
{\em Phys. Rev. D\/} {\bf 57} (1998)   4724.


\bibitem{kk1998}
K.\ Konno, M.\ Kasai,
{\em Prog. Theor. Phys.\/}
{\bf 100} (1998) 1145.


\bibitem{jw2001}  
J.\ Wudka,
{\em Phys. Rev. D\/} {\bf 64} (2001)   065009.

\bibitem{gl}  
S. Capozziello, G. Lambiase,
{\em Mod. Phys. Lett. A\/} {\bf  14} (1999)   2193.


\bibitem{ac}  
A.\ Camacho,
{\em Mod. Phys. Lett. A\/} {\bf 14} (1999) 2545.


\bibitem{cpsc}
A.\ Cho, in News of the Week, 
{\em Science \/} {\bf 287} (2000) 2395.

\bibitem{gac2}
G.\ Amelino-Camelia,
{\em Nature\/} {\bf 398} (1999) 216.

\bibitem{dva_nv}
D.\ V.\ Ahluwalia,
{\em Nature\/} {\bf 398} (1999) 199.

\bibitem{grf2001}  
D.\ V.\ Ahluwalia, M.\ Kirchbach,
{\em Int. J. Mod. Phys. D\/} {\bf 10} (2001)   811.


\bibitem{QSO1}  
J.\ K.\ Webb {\em et al.\/},
{\em Phys. Rev. Lett.\/} {\bf 82} (1999)   884.


\bibitem{QSO2}
J.\ K.\ Webb {\em et al.\/},
{\em Phys. Rev. Lett.\/} {\bf 87} (2001)  091301.  

\bibitem{QSO3}
M.\ T.\ Murphy {\em et al.\/} 
{\em Mon. Not. Roy. Astron. Soc.\/} {\bf 327} (2001) 1223.

\end{chapthebibliography}
\end{document}